\documentstyle[11pt,twoside,paspconf]{article}

\begin{document}

\title{A New Look at Poor Groups of Galaxies}

\author{Ann I. Zabludoff}
\affil{UCO/Lick Observatory and Astronomy Department, University of California,
    Santa Cruz, CA 95064}

\begin{abstract}
We use a new fiber spectroscopic survey of 12 nearby, poor groups of
galaxies to examine the dynamics and evolution of galaxies in these
common, but poorly studied, environments.  Some of our conclusions
are: (1) The nine groups in our sample with diffuse X-ray emission are
in fact bound systems with at least 20-50 group members with absolute
magnitudes as faint as $M_B \sim -14 + 5\log_{10}h$.  (2) Galaxies in
each X-ray-detected group have not all merged together because a
significant fraction of the group mass lies outside of the galaxies
and in a common halo, thereby reducing the rate of galaxy-galaxy
interactions.  (3) The similarity of the recent star formation
histories and the fraction of early type galaxies in some poor groups
to those in rich clusters suggests that cluster-specific environmental
effects may not play a dominant role in the recent evolution of
cluster galaxies.  The evolution of group and cluster members may be
driven instead by galaxy-galaxy interactions, which are likely to
occur with equal frequency in field groups and in groups that have
recently fallen into clusters (i.e., subclusters).

\end{abstract}




\section{Introduction}
Most galaxies in the nearby universe, including our own Galaxy, belong
to poor groups of galaxies.  Despite the ubiquity of the group
environment, we know little about the internal dynamics of groups and
the evolution of group galaxies outside of the Local Group.  To learn
whether the diverse and detailed results on the Local Group presented
at this meeting apply elsewhere, we must also survey other poor
groups.

Because poor groups are apparent systems of fewer than five bright
($\leq M^*$) galaxies, past studies have been hampered by small number
statistics.  Some of the critical, unanswered questions are (1) whether
most poor groups are real systems instead of chance projections of
galaxies along the line-of-sight, (2) why many poor groups, with their
favorable environments for galaxy-galaxy mergers, have survived until
now, (3) and how galaxies evolve in groups, where the collisional
effects of the intragroup gas and the tidal influences of the global
potential are weaker than in rich clusters.  The advent of
multi-object spectroscopy now makes it possible to address such
questions in unprecedented detail.  In this talk, I discuss the first
results obtained in collaboration with John Mulchaey (Carnegie
Observatories) from a fiber spectroscopic survey of 12 nearby, poor
groups (Zabludoff \& Mulchaey 1998a; Mulchaey \& Zabludoff 1998;
Zabludoff \& Mulchaey 1998b).

\section{Poor Groups:  What are They?}

\subsection{Three Classes of Groups}

Poor groups in the literature have been identified optically and fall
into several classes that can be distinguished by their X-ray
properties and bright galaxy morphologies.  Groups with detectable, hot
intra-group gas typically have a giant ($\leq M^* - 1$) elliptical
that is the brightest group galaxy (BGG) and that lies near or on the
peak of the smooth, symmetric X-ray 
emission (Mulchaey et al.\ 1996).  In contrast,
groups without extended X-ray emission tend to optically resemble the
Local Group, which consists of a few bright, late-type galaxies and
their satellites.  If some non-X-ray-detected, late-type-dominated
groups evolve into groups with a central, giant elliptical and a
detectable intra-group medium, then groups in transition may form a
third class of objects.  We would expect the cores of such systems to
have signatures of recent dynamical evolution, including interacting
galaxies and X-ray gas that does not coincide with the galaxies.  To
construct our sample, we select poor groups from these three classes 
that also have complementary X-ray images.

The sample consists of 12 nearby ($1500 < cz < 8000$ km\,s$^{-1}$),
optically-selected groups from the literature for which there are
existing, sometimes serendipitous, pointed ROSAT/PSPC X-ray
observations:  NGC\,533, NGC\,5129, NGC\,5946, NGC\,4325, NGC\,741, NGC\,2563,
HCG\,42, HCG\,62, HCG\,90, NGC 664, NCG 491, and NGC\,7582. 
Nine of the groups are X-ray-detected.  HCG\,90 (Hickson 1982), a
possible transitional object with several interacting galaxies in its
core (Longo et al.\ 1994), is only marginally X-ray-detected
(Ponman et al.\  1996).  It is important to stress that our group
sample is biased with respect to published group catalogs, in which
less than half of the groups are X-ray-detected.  The sample
is weighted toward X-ray groups, because of the likelihood that they
are bound systems and the most evolved poor groups.

\subsection{Bound or Superpositions?}

The issue of whether many poor groups, even those identified from
redshift surveys, are bound systems instead of chance superpositions
of galaxies along the line-of-sight has been a puzzle.  The existence
of one poor group, our Local Group, is unchallenged.  In contrast,
Ramella et al.\ (1989) show that $\sim 30\%$ of groups of three or
four galaxies in the CfA Redshift Survey (Huchra et al.\ 1995) are
probably unbound, geometric projections.  The detection of a
significant population of fainter members would be an important first
step in identifying poor groups that are real.

Using the Las Campanas fiber spectrograph designed by Steve Shectman,
we measured the $\sim 100$ brightest galaxies projected within $1.5
\times 1.5$ deg$^2$ of the center of each of the 12 sample groups.
The radial velocity distributions for galaxies in the fields of the
nine X-ray-detected groups each reveal $\sim 20$-50 group members down
to absolute magnitudes of $M_B \sim -14$ to $-16 + 5\log_{10}h$.
The surprisingly large membership, the central concentration of early
type galaxies, the similarity of the BGG's position and orientation to
those of the diffuse X-ray emission, the consistency of the optical
velocity dispersion and the X-ray temperature, and the short crossing
times ($\leq 0.05$ of a Hubble time) of the X-ray groups suggest that
they are bound systems, not geometric superpositions of galaxies, and
that the group cores are close to virialization or virialized.

Because we do not detect diffuse X-ray emission and find only 5-8
members in the three non-X-ray-detected groups, we are unable to
determine their dynamical state.  The non-X-ray groups, which consist
of one or two $M^*$ or brighter spirals with several fainter galaxies
that may be satellites, are {\it morphologically} akin to the Local Group
(although our samples are not sufficiently deep to ascertain whether
any group has a dwarf spheroidal population like that of the Local
Group; van den Bergh 1992).  If the non-X-ray groups are {\it dynamically}
similar to the Local Group, they are bound (see Zaritsky 1994).  Our
current data do not exclude this possibility --- the velocity
dispersions of the non-X-ray groups are consistent with the upper
limits on their X-ray luminosities (Mulchaey \& Zabludoff 1998).

\section{Poor Groups:  Why are They?}

If some poor groups are bound systems, then another critical question
is why they exist at all.  Poor groups have higher galaxy densities
than the field and lower velocity dispersions than cluster cores,
making them favorable sites for galaxy-galaxy mergers (Barnes 1985).
The likelihood of mergers and the short group crossing times ($\leq
0.05$ of a Hubble time) suggest that most groups should have already
merged into one object.  Therefore, either bound groups are collapsing
for the first time like the Local Group (Zaritsky 1994) or only a
small fraction of the group mass is tied to the galaxies, lowering the
rate of galaxy-galaxy interactions relative to a galaxy-dominated
system and allowing the group to survive many crossing times
(Governato et al.\ 1991; Bode et al.\ 1993; Athanassoula et al.\ 1997).  To
address this issue by measuring the underlying mass distribution of
poor groups, we use the improved statistics of our spectroscopic
survey.

The velocity dispersion of the combined group sample does
not decrease significantly with radius from the central $\sim
0.1$$h^{-1}$ Mpc to at least $\sim 0.5$$h^{-1}$ Mpc, in contrast to
the more than factor of two decrease that would be observed if the
entire mass were concentrated within 0.1$h^{-1}$ Mpc.  The extended
mass is either in the galaxies, in a common halo through which the
galaxies move, or in both the galaxies and a diffuse halo.  If all the
mass were tied to the galaxies, most of the mass would be associated
with the bright, central elliptical in those groups in which the BGG
dominates the light.  For groups with a few galaxies that have
luminosities comparable to the BGG, the velocity dispersion would be
increased at large radii by subgroups consisting of a massive galaxy
and the subgroup members that are orbiting it.  If this picture were
accurate, we would expect the velocity dispersion profiles of groups
with several comparably bright galaxies to be shallower than those in
which the BGG is dominant.  However, the combined velocity dispersion
profile of a subsample of two groups (HCG\,42 and NGC\,741) in which the
BGG dominates the light (i.e., the BGG luminosity exceeds the
combined luminosity of the other $M^*$ or brighter galaxies) is
indistinguishable from that of the entire sample.  Therefore, we
conclude that most of the group mass lies in a smooth, extended dark
halo.  

This result argues that poor groups survive longer than predicted by
models in which all the mass is tied to individual galaxies and may
explain why so many poor groups are observed in lieu of single merger
remnants.

\section{Poor Groups:  How do Galaxies Evolve in Them?}

\subsection{Distribution of Early Type Fractions}

The factors that might affect the evolution of galaxies in poor groups
are different from those present in rich clusters.
If clusters evolve hierarchically by accreting
poor groups of galaxies (subclusters), members of an infalling group
have recently experienced the hot, dense cluster environment 
for the first time.
Therefore, galaxies in poor groups in the field
are a control sample for understanding the factors that
influence the evolution of their counterparts in subclusters.

For example, we can compare the
morphologies and recent star formation histories of
galaxies in the subclusters of complex clusters like Coma
(Caldwell \& Rose 1997) with those of galaxies in
poor field groups.  Differences between the samples would argue
that cluster environment is important in transforming galaxies
at the present epoch.
On the other hand, the lack of such differences
would suggest, as the simplest explanation,
either that star formation and morphology are influenced by
mechanisms present in both field groups and subclusters, such as galaxy-galaxy
encounters, or that the effects of environment on
galaxies are insignificant compared
with conditions at the time of galaxy formation.

Despite the usefulness of group galaxies 
as a control sample, their properties, especially at faint magnitudes,
are not well-known.
Past work has included only the four
or five brightest galaxies, which biases samples toward ellipticals,
and has targeted only the central $\leq 0.3$$h^{-1}$ Mpc,
where early types concentrate.  Therefore,
to compare the morphologies and star formation histories of group 
and cluster members, we must sample both environments to similar
physical radii and absolute magnitude limits.

In the X-ray groups, the early type fraction ($f$) 
ranges widely from $\sim 55\%$
(HCG\,62, NGC\,741, and NGC\,533) 
to $\sim 25\%$ (e.g., NGC\,2563).  The latter value 
is characteristic of the field ($\sim 30\%$; Oemler 1992).
We find no early types among the 6-8 galaxies in each of
the three non-X-ray groups.
The early type fractions of $\sim 55\%$ in 
NGC\,533, NGC\,741, and HCG\,62 are most surprising, because they are consistent with
those of rich clusters for similar
physical radii and absolute magnitude limits ($\sim 0.55$-0.65; 
Whitmore et al.\ 1993).

\subsection{Early Type Fraction vs. Velocity Dispersion}

The correlation between early type fraction and group
velocity dispersion is significant at the $>0.999$ level.
The form of the relation cannot be the same for rich clusters ---
our fit to the group data predicts that the early type fraction
in a $\sigma_r \sim 1000$ km\,s$^{-1}$ cluster is an unphysical $f = 124\%$!
Therefore, the relation must turn up between the poor group and rich
cluster regimes.  The group $f - \sigma_r$ relation implies either
that galaxy morphology is set by the local potential size at the
time of galaxy formation (Hickson, Huchra, \& Kindl 1988) and/or that 
$\sigma_r$ and $f$ increase as a group 
evolves (Diaferio et al.\ 1995).

The early type fractions in our highest velocity dispersion
groups ($\sigma_r \sim 400$ km\,s$^{-1}$) are as high as in rich clusters.
If some early type galaxies are evolved merger remnants, 
then the galaxy populations of
higher velocity dispersion groups are more evolved on average.
At some point in the group's evolution, perhaps at a
velocity dispersion near 400 km\,s$^{-1}$, any morphological
evolution resulting from galaxy mergers ceases,
and the fraction of merger remnants in poor groups and rich clusters
is comparable.  The implicit upturn in our $f - \sigma_r$ relation
suggests such a saturation point.

Alternatively, the similarity of some group and cluster
early type fractions, and the steepening of 
the $f - \sigma_r$ relationship at high $\sigma_r$,
might arise from conditions at the time of galaxy formation.  For example,
it is possible that poor groups such as NGC\,533 and rich clusters like
Coma begin as similar mass density perturbations
with correspondingly similar
galaxy populations.  In this simple picture, NGC\,533 does not develop a
cluster-size potential, because its field lacks the surrounding
groups that Coma later accretes.  

In summary, the cluster-like fraction of early type galaxies
in NGC\,533, NGC\,741, and HCG\,62 indicate that clusters are not the only
environments with copious quantities of E and S0 galaxies.
The simplest explanation is either that fluctuations in
the initial conditions permitted early types to form
in these groups' comparatively low velocity dispersion, low galaxy density
environments or that the galaxies' subsequent evolution
was the product of a mechanism,
such as galaxy-galaxy interactions,
common to both groups in the field and groups that become subclusters.
Although additional environmental 
mechanisms may affect the evolution of cluster galaxies, such
cluster-specific processes are not required to explain the current data.
A cluster that evolves hierarchically from subclusters with the properties
of NGC\,533, NGC\,741, and HCG\,62 will have, at least initially, a similar
galaxy population.  

\subsection{Star Formation in Early Types}

The star formation histories of galaxies in poor groups provide
additional insight into the environmental factors that may influence
the evolution of galaxies.  One approach is to examine the spectra of
the early types for evidence of on-going star formation or of a young
stellar population.  We can then compare the fraction of E and S0
group members that have recently formed stars with a sample from rich
clusters with complex structure.

Star formation is on-going or has ended within the last $\sim 2$ Gyr
in eight of the 64 early type group members for which we have
spectra ($12\%$).  This fraction is roughly the same for clusters, such as
Coma, with infalling groups ($\sim 15\%$; Caldwell \& Rose 1997).
This result, and the similarity of the early type fractions of some
poor groups to rich clusters, suggests that an environmental mechanism
present in both groups and subclusters may be responsible.  For
example, galaxy-galaxy encounters can produce bursts of star formation
(e.g., Londsdale et al.\ 1984; Kennicutt et al.\ 1987;
Sanders {et al.\ 1988) in which the gas is consumed or stripped
away.

Although our observations of poor groups suggest that the effects of
cluster environment are not necessary to produce the early type
fractions and star formation episodes of nearby clusters, we suspect
that the star formation histories of group and subcluster galaxies
will begin to deviate after the subcluster and cluster mix.  Proposed
gas removal processes including ram pressure stripping and the tidal
limitation of galaxy halos, which are more
efficient in clusters than in groups, may eventually suppress star
formation in cluster galaxies.  Comparative studies of the H{\sc i} content
of field, group, and cluster galaxies will help to resolve this issue.

\acknowledgments
This research was supported by grants from the NSF and NASA.


%
%

\begin{references}

\reference Athanssoula, E., Makino, J., \& Bosma, A. 1997, \mnras, 286, 825
\reference Barnes, J. 1985, \mnras, 215, 517
\reference Bode, P.W., Cohn, H.N.,  Lugger, P.M. 1993, \apj, 416, 17
\reference Caldwell, N., Rose, J. 1997, \aj, 113, 492
\reference Diaferio, A., Geller, M., Ramella, M. 1995, \aj, 109, 2293 
\reference Governato, F., Bhatia, R., Chincarini, G. 1991, \apjlett,
371, L15
\reference Hickson, P., Huchra, J.,  Kindl, E. 1988, \apj, 331, 64
\reference Hickson, P. 1982, \apj, 255, 382
\reference Huchra, J.P., Geller, M.J.,  Corwin, H. 1995, \apjsupp, 99, 391
\reference Kennicutt, R.C., Roettiger, K.A., Keel, W.C., Van der Hulst,
J.M., Hummel, E. 1987, \apj, 93, 1011,
\reference Lonsdale, C.J., Persson, S.E., Matthews, K. 1984, \apj, 287, 95
\reference Longo, G., Busarello, G., Lorrenz, H., Richter, G., Zaggia, S. 1994, \aap, 282, 418
\reference Mulchaey, J.S., Zabludoff, A.I. 1998, \apj, 496, 73
\reference Mulchaey, J., Davis, D., Mushotzky, R., Burstein, D. 1996, \apj, 456, 80
\reference Oemler, A. 1992, in: {\it Clusters and Superclusters of Galaxies},
(ed.) A.C.\ Fabian (Dordrecht: Kluwer), p.~29
\reference Ponman, T., Bourner, P., Ebeling, H., \& Bohringer, H. 1996, 
\mnras, 283, 690
\reference Ramella, M., Geller, M.J., Huchra, J.P. 1989, \apj, 344, 57
\reference Sanders, D.B., Soifer, B.T., Elias, J.H., Matthews, K., 
Madore B.F. 1988, \apj, 325, 74
\reference van den Bergh, S. 1992, \aap, 264, 75
\reference Whitmore, B., Gilmore, D., Jones, C. 1993, \apj, 407, 489
\reference Zabludoff, A.I., Mulchaey, J.S. 1998a, \apj, 496, 39
\reference Zabludoff, A.I., Mulchaey, J.S. 1998b, \apjlett, 498, L5
\reference Zaritsky, D. 1994, in:  {\it The Local Group:  Comparative and Global
Properties}, (eds.) A.\ Layden, R.C.\ Smith, \& J.\ Storm, ESO Conference and
Workshop Proceedings No.\ 51, p.~187
\end {references}
\section*{Discussion}
{\it Terndrup:}  Could you please elaborate on your remarks about the 
coupling between the elliptical fraction in groups and group environment?\\

\noindent {\it Zabludoff:}  We observe a strong correlation between group
early-type fraction and velocity dispersion.  This correlation could result
either from an increase in the early type fraction and velocity dispersion
as a group evolves, where galaxy morphologies change due to a mechanism such
as galaxy-galaxy mergers, or from conditions at the time of galaxy formation
that tie the E and S0 fraction to the potential well depth early on.  One
point suggesting that group galaxy evolution is affected by environment at
later epochs is that this correlation has a more shallow slope than that for
rich clusters.  In other words, there seems to be a saturation point at
which early type fraction stops increasing with velocity dispersion.  This
point is at about 400 to 500 $\rm km\,s^{-1}$, which is about the velocity
dispersion that a poor group would require such that an $M^*$ galaxy would
experience a merger within a Hubble time.  Therefore, it is possible that
mergers cause some evolution in the early type fraction of poor groups and
cease to be effective in richer groups and clusters.
\end {document}